\documentclass[10pt,pra,twocolumn,showpacs]{revtex4}
\usepackage{amsmath}
\usepackage[dvips]{graphicx}
\usepackage{amsfonts}
\usepackage{float}
\usepackage{amssymb}

\begin{document}

\title{Applying Gaussian quantum discord to quantum key distribution}
\author{Xiaolong Su}
\email{suxl@sxu.edu.cn}
\affiliation{State Key Laboratory of Quantum Optics and Quantum Optics Devices, Institute
of Opto-Electronics, Shanxi University, Taiyuan, 030006, People's Republic
of China}

\begin{abstract}
In this paper, we theoretically prove that the Gaussian quantum discord
state of optical field can be used to complete continuous variable (CV)
quantum key distribution (QKD). The calculation shows that secret key can be
distilled with a Gaussian quantum discord state against entangling cloner
attack. Secret key rate is increased with the increasing of quantum discord
for CV QKD with the Gaussian quantum discord state. Although the calculated
results point out that secret key rate using the Gaussian quantum discord
state is lower than that using squeezed state and coherent state at the same
energy level, we demonstrate that the Gaussian quantum discord, which only
involving quantum correlation without the existence of entanglement, may
provide a new resource for realizing CV QKD.
\end{abstract}

\maketitle

\section{Introduction}

Quantum correlation, which is measured by quantum discord \cite%
{Ollivier2001,Modi,Aar}, is a fundamental resource for quantum information
processing tasks. It has been shown that some quantum computational tasks
based on a single qubit can be carried out by separable (that is,
non-entangled) states that nonetheless carries non-classical correlations 
\cite{Knill1998,Ryan2005,Lanyon2008}. Recently, quantum discord is extended
to two-mode Gaussian states \cite{Giorda2010,Adesso2010}. A two-mode
Gaussian state is entangled with Gaussian quantum discord $D>1$, when $0\leq
D\leq 1$ we have either separable or entangled states. Gaussian quantum
discord has been experimentally demonstrated too \cite%
{Gu2012,Blandino2012,Madsen2012}.

Quantum key distribution (QKD) allows two legitimate parties, Alice and Bob
who are linked by a quantum channel and an authenticated classical channel,
to establish the secret key only known by themselves. Continuous variables
(CV) QKD using Gaussian quantum resource state, such as entangled state,
squeezed state and coherent state, as the resource state, along with
reconciliation and privacy amplification procedure to distill the secret key 
\cite{Weedbrook2012}. There are two type of QKD schemes, one is called
prepare-and-measure scheme, the other is entanglement-based scheme. The
equivalence between these two type CV QKD schemes has been proved. QKD with
coherent state (squeezed state) has been proved to be equivalent to
heterodyning (homodyning) one of the two entangled modes of an
Einstein-Podolsky-Rosen (EPR) entangled state \cite{Gros1}. Generally, the
entanglement-based QKD model is used to investigate the security of CV QKD.
The security of CV QKD scheme has been analyzed \cite{Ibl,Gros2,Nav1}, and
it has been proved to be unconditionally secure, that is, secure against
arbitrary attacks over long distance \cite{Ren,Lev}. Recently, a CV QKD
scheme with thermal states is also proposed and proved to be secure against
collective Gaussian attacks \cite{Weedbrook2}.

Very recently, it has been shown that quantum discord can be used as a
resource for QKD in general \cite{Pir}. What we concerned is the role of
Gaussian quantum discord in CV QKD. In this paper, we apply a two-mode
Gaussian discord state, where only quantum correlation exists and without
entanglement, to implement CV QKD. The calculation shows that the secret key
can be distilled with the two-mode Gaussian discord state against entangling
cloner attack, which is the most important and practical example of
collective Gaussian attack. The secret key rate of the QKD scheme with
Gaussian discord state is increased with the increasing of the quantum
discord. The secret key rates of the CV QKD schemes with the Gaussian
discord state, squeezed state and coherent state (no-switching QKD) are
compared. Although squeezed state and coherent state offer higher secret key
rate than the Gaussian discord state, we demonstrate the Gaussian discord
can be used to establish secret key.

\section{The Gaussian discord state and QKD scheme}

The QKD scheme with a two-mode Gaussian quantum discord state and entangled
state is shown in Fig. 1. Figure 1(a) shows a two-mode Gaussian discord
state, as shown in \cite{Gu2012}, which is prepared by correlated
(anti-correlated) displacement of two coherent states in the amplitude
(phase) quadrature with a discording noise $V$. Figure 1(b) shows an EPR
entangled state with a variance $V_{E}=\cosh 2r$, where $r\in \lbrack
0,\infty )$ is the squeezing parameter. The amplitude and phase quadratures
of an optical mode $\hat{a}$ are defined as $\hat{X}_{a}=\hat{a}+\hat{a}%
^{\dagger }$ and $\hat{Y}_{a}=(\hat{a}-\hat{a}^{\dagger })/i$, respectively.
The variances of amplitude and phase quadratures for a vacuum (coherent)
state are $V(\hat{X}_{v})=V(\hat{Y}_{v})=1$. The covariance matrix of the
two-mode Gaussian quantum resource state in Fig. 1(a) and (b) is given by 
\begin{equation}
\mathbf{\sigma }=\left( 
\begin{array}{cc}
\alpha \mathbf{I} & \gamma \mathbf{Z} \\ 
\gamma \mathbf{Z} & \beta \mathbf{I}%
\end{array}%
\right) ,
\end{equation}%
where $\mathbf{I}$ and $\mathbf{Z}$ are the Pauli matrices 
\begin{equation}
\mathbf{I=}\left( 
\begin{array}{cc}
1 & 0 \\ 
0 & 1%
\end{array}%
\right) ,\qquad \mathbf{Z}=\left( 
\begin{array}{cc}
1 & 0 \\ 
0 & -1%
\end{array}%
\right) ,
\end{equation}%
$\alpha =\beta =V_{D}=V+1$, $\gamma =V$ for the two-mode Gaussian discord
state and $\alpha =\beta =V_{E}$, $\gamma =$ $\sqrt{V_{E}^{2}-1}$ for the
EPR entangled state, respectively.

\begin{figure}[tbp]
\setlength{\belowcaptionskip}{-3pt} 
\centerline{
\includegraphics[width=80mm]{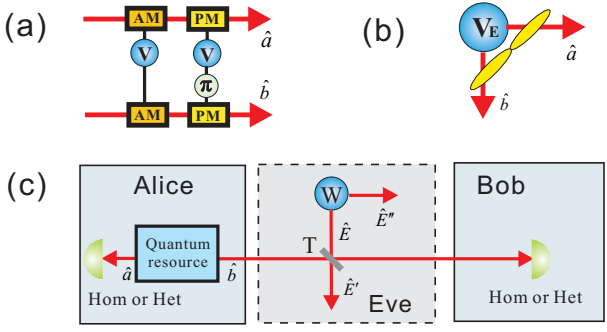}
}
\caption{Schematic of the CV QKD scheme with a two-mode Gaussian state. (a):
The two-mode Gaussian discord state, AM: amplitude modulator, PM: phase
modulator, $\protect\pi $: $\protect\pi $ phase shift. (b): EPR entangled
state, (c): the CV QKD scheme. The transmission efficiency of quantum
channel is modeled by a beam splitter with transmission T. Eve performs
entangling cloner attack, where the variance of the ancillary EPR state is
W. Hom: homodyne detection, Het: heterodyne detection.}
\label{Fig1}
\end{figure}

Quantum discord is defined as the difference between two quantum analogues
of classically equivalent expression of the mutual information. The Gaussian
quantum discord of a two-mode Gaussian state is given by \cite{Adesso2010}

\begin{equation}
D_{AB}=f(\sqrt{I_{2}})-f(\nu _{-})-f(\nu _{+})+f(\sqrt{E^{\min }}),
\label{discord}
\end{equation}%
where $f(x)=(\frac{x+1}{2})\log {\frac{x+1}{2}}-(\frac{x-1}{2})\log {\frac{%
x-1}{2}}${, }%
\begin{equation}
\nu _{\pm }=\sqrt{\frac{\Delta \pm \sqrt{\Delta ^{2}-4\det \mathbf{\sigma }}%
}{2}}
\end{equation}%
are the symplectic eigenvalues of a two-mode covariance matrix $\mathbf{%
\sigma }=\left( 
\begin{array}{cc}
\mathbf{A} & \mathbf{C} \\ 
\mathbf{C} & \mathbf{B}%
\end{array}%
\right) ${\ with }$\det \mathbf{\sigma }$ as the determinant of covariance
matrix and $\Delta =\det \mathbf{A}+\det \mathbf{B}+2\det \mathbf{C}$, and

\begin{equation}
E^{\min }=\left\{ 
\begin{array}{ll}
\frac{2I_{3}^{2}+(I_{2}-1)(I_{4}-I_{1})+2|I_{3}|\sqrt{%
I_{3}^{2}+(I_{2}-1)(I_{4}-I_{1})}}{(I_{2}-1)^{2}}\quad  & a) \\ 
\frac{I_{1}I_{2}-I_{3}^{2}+I_{4}-\sqrt{%
I_{3}^{4}+(I_{4}-I_{1}I_{2})^{2}-2I_{3}^{2}(I_{4}+I_{1}I_{2})}}{2I_{2}}\quad 
& b)%
\end{array}%
\right. 
\end{equation}%
where a) applies if $\quad (I_{4}-I_{1}I_{2})^{2}\leq
I_{3}^{2}(I_{2}+1)(I_{1}+I_{4})$ and b) applies otherwise. $I_{1}=\det 
\mathbf{A}$, $I_{2}=\det \mathbf{B}$, $I_{3}=\det \mathbf{C}$, $I_{4}=\det {%
\sigma }$ are the symplectic invariants.

PPT criterion is a necessary and sufficient criterion for entanglement of
Gaussian state \cite{Simon,Werner}. A Gaussian state is entangled iff $%
\tilde{\nu}_{-}<1$, where $\tilde{\nu}_{-}$ is the smallest symplectic
eigenvalue of partial transposed covariance matrix for two-mode Gaussian
state, which is given by \cite{Serafini2004}

\begin{equation}
\tilde{\nu}_{-}=\sqrt{\frac{\tilde{\Delta}-\sqrt{\tilde{\Delta}^{2}-4\det 
\mathbf{\sigma }}}{2}}
\end{equation}%
where $\tilde{\Delta}=\det \mathbf{A}+\det \mathbf{B}-2\det \mathbf{C}$.

Based on the covariance matrix in eq. (1) for the Gaussian discord state, we
calculated the quantum discord and smallest symplectic eigenvalue of PPT
criterion, which are shown in Fig. 2. As shown in Fig. 2(a), the quantum
discord is increased dramatically with the increasing of input variance $%
V_{D}$ in the region of $V_{D}$ $\in \lbrack 1,100]$. When $V_{D}>100$, the
quantum discord increased slowly with the increasing of $V_{D}$. The
smallest quantum discord is 0.12 at $V_{D}=1$. The quantum discord is always
smaller than 1. In Fig. 2(b), the smallest symplectic eigenvalue of partial
transposed covariance matrix is always 1, which means that there is no
entanglement in the Gaussian discord state.

\begin{figure}[tp]
\setlength{\belowcaptionskip}{-3pt} 
\centerline{
\includegraphics[width=80mm]{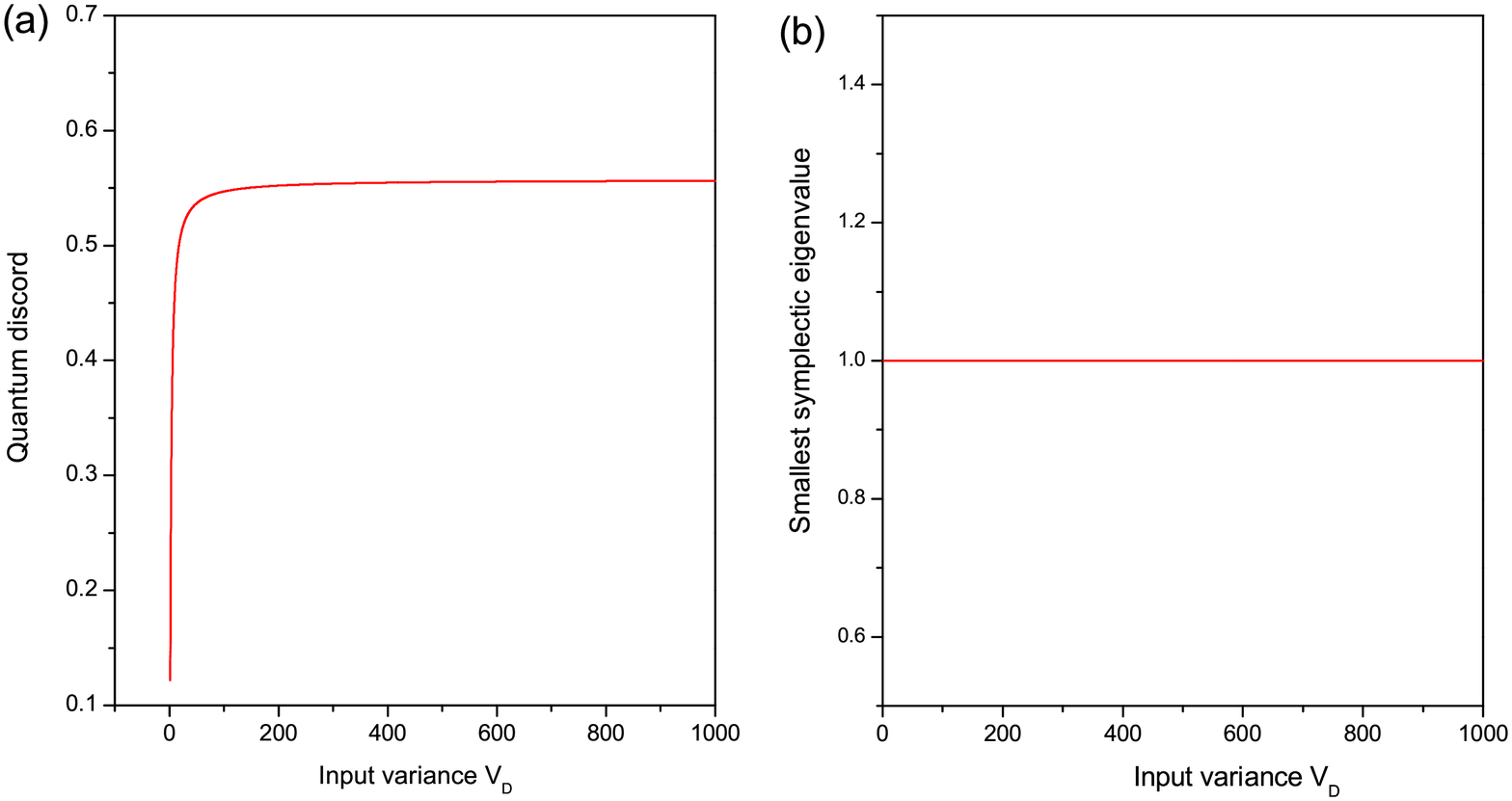}
}
\caption{Quantum discord (a) and smallest symplectic eigenvalue of PPT
criterion (b) for the Gaussian discord state. }
\label{Fig2}
\end{figure}

Figure 1(c) shows the CV QKD scheme with a two-mode Gaussian state as
quantum resource state, which can be the two-mode Gaussian discord state or
the EPR entangled state. Alice hold mode $\hat{a}$, and transmitted mode $%
\hat{b}$ to Bob over the quantum channel. Here, we consider that Alice and
Bob perform homodyne (Hom) or heterodyne (Het) detection on their own beam,
which corresponds to the CV QKD scheme with homodyne or heterodyne
detection. We assume that Eve perform entangling cloner attack \cite{Gros1},
which is the most important and practical example of a collective Gaussian
attack \cite{Ren,Nav,Gar,Pira}, to steal the information. She prepares an
ancillary EPR entangled states with variance $W$, which corresponds to the
excess noise $\delta =W-1$ in \cite{Nam} and $\epsilon =(W-1)(1-T)/T$ in 
\cite{Gros1}. $W=1$ means there is no excess noise ($\delta =0$) in the
channel, when $W>1$, there is excess noise ($\delta =W-1$) in the channel.
She keeps one mode $\hat{E}^{\prime \prime }$ and mixed the other mode $\hat{%
E}$ with the transmitted mode $\hat{b}$ in the quantum channel by a beam
splitter, leading to the output mode $\hat{E}^{\prime }$. Eve's output modes
are stored in a quantum memory and detected collectively at the end of the
protocol. Eve's final measurement is optimized based on Alice and Bob's
classical communication. After communication is completed, Alice and Bob
perform reconciliation, error correction \cite{Bra, Bennett} and privacy
amplification \cite{Ca} to distill final secret key.

\section{Security of the CV QKD scheme}

\subsection{Homodyne detection}

In the CV QKD scheme with homodyne detection, Alice and Bob perform homodyne
detection on their own beams to measure the amplitude or phase quadrature,
respectively. For CV QKD with EPR entangled state, homodyning one of the
entangled beam is equivalent to the CV QKD with squeezed state. So we will
compare the Gaussian discord state QKD with squeezed state QKD in this
section. In the following, we use the variable $X$ to represent amplitude or
phase quadrature of an optical mode to analyze the secret key without losing
the generality.

\textit{Direct reconciliation}. In direct reconciliation, Bob attempts to
guess what Alice sent. The secret key rate is given by 
\begin{equation}
K_{DR}=I(X_{A}\colon X_{B})-I(X_{A}\colon E),
\end{equation}%
where%
\begin{equation}
I(X_{A}\colon X_{B})=H(X_{B})-H(X_{B}|X_{A}),
\end{equation}%
is the mutual information between Alice and Bob, with $H(X_{B})=(1/2)\log
_{2}V(X_{B})$ and $H(X_{B}|X_{A})=$ $(1/2)\log _{2}V(X_{B}|X_{A})$ being the
total and conditional Shannon entropies. Eve's information is 
\begin{equation}
I(X_{A}\colon E)=S(E)-S(E|X_{A}),
\end{equation}%
where $S(\cdot )$ is the von Neumann entropy. The von Neumann entropy of a
Gaussian state $\rho $ can be expressed in terms of its symplectic
eigenvalues \cite{Holevo} 
\begin{equation}
S(\rho )=\sum\limits_{k=1}^{n}g(\nu _{k}),
\end{equation}%
with $g(\nu )=\frac{1}{2}\left( \nu +1\right) \log _{2}[\frac{1}{2}\left(
\nu +1\right) ]-\frac{1}{2}\left( \nu -1\right) \allowbreak \log _{2}[\frac{1%
}{2}\left( \nu -1\right) ]$, where $\mathbf{\nu }$ $=\{\nu _{1},...\nu
_{n}\} $ are the symplectic eigenvalues of Gaussian state $\rho $. The
symplectic spectrum $\mathbf{\nu }$ $=\{\nu _{1},...\nu _{n}\}$ of an
arbitrary correlation matrix $\mathbf{\sigma }$ can be calculated by finding
the (standard) eigenvalues of the matrix $\left\vert i\Omega \mathbf{\sigma }%
\right\vert $, where $\Omega $ defines the symplectic form and is given by 
\cite{Weedbrook2012} 
\begin{equation}
\Omega =\bigoplus\limits_{k=1}^{n}\left( 
\begin{array}{cc}
0 & 1 \\ 
-1 & 0%
\end{array}%
\right) .
\end{equation}%
Here $\bigoplus $ is the direct sum indicating adding matrices on the block
diagonal.

In Fig. 1(c), the covariance matrix of the two-mode Gaussian state
distributed between Alice and Bob in the CV QKD is given by 
\begin{equation}
\mathbf{\sigma }_{AB}=\left( 
\begin{array}{cc}
V_{A}\mathbf{I} & \gamma ^{\prime }\mathbf{Z} \\ 
\gamma ^{\prime }\mathbf{Z} & V_{B}\mathbf{I}%
\end{array}%
\right) ,
\end{equation}%
where $V_{A}=V_{E}$, $V_{B}=TV_{E}+(1-T)W$, $\gamma ^{\prime }=\sqrt{%
T(V_{E}^{2}-1)}$ for the EPR entangled state and $V_{A}=V_{D}$, $%
V_{B}=TV_{D}+(1-T)W$, $\gamma ^{\prime }=\sqrt{T}V$ for the Gaussian discord
state, respectively.

The conditional variance is defined as \cite{Grangier} $V_{X\mid
Y}=V(X)-\left\vert \left\langle XY\right\rangle \right\vert ^{2}/V(Y)$. So
Bob's conditional variance in homodyne detection is given by 
\begin{equation}
V_{B\mid A}=V_{B}-\frac{\gamma ^{\prime 2}}{V_{A}}.
\end{equation}%
The mutual information between Alice and Bob is $I^{Hom}(X_{A}\colon X_{B})=%
\frac{1}{2}\log _{2}[V_{B}/V_{B\mid A}]$, which is same for the direct and
reverse reconciliation.

Eve's covariance matrix is made up from the modes $\hat{E}^{\prime }$ and $%
\hat{E}^{\prime \prime }$, which is 
\begin{equation}
\mathbf{\sigma }_{E}=\left( 
\begin{array}{cc}
e_{v}\mathbf{I} & \varphi \mathbf{Z} \\ 
\varphi \mathbf{Z} & W\mathbf{I}%
\end{array}%
\right),
\end{equation}%
where $e_{v}=(1-T)V_{A}+TW$, $\varphi =\sqrt{T(W^{2}-1)}$.

In order to obtain $S(E|X_{A})$ we need to calculate the symplectic spectrum
of the conditional covariance matrix $\mathbf{\sigma }_{E\mid X_{A}}$, which
represents the covariance matrix of Eve's system where mode $\hat{a}$ has
been measured by Alice using homodyne detection and is given by \cite%
{Weedbrook2012,Eisert2002,Fiu2002} 
\begin{equation}
\mathbf{\sigma }_{E|X_{A}}=\mathbf{\sigma }_{E}-(V_{A})^{-1}\mathbf{D}\Pi%
\mathbf{D}^{T},
\end{equation}%
where 
\begin{equation}
\Pi =\left( 
\begin{array}{cc}
1 & 0 \\ 
0 & 0%
\end{array}%
\right),
\end{equation}%
and $\mathbf{D}$ is the matrix describing the quantum correlations between
Eve' modes and Alice's mode, which is given by 
\begin{equation}
\mathbf{D=}\left( 
\begin{array}{c}
\left\langle X_{E%
%TCIMACRO{\U{b4}}%
%BeginExpansion
{\acute{}}%
%EndExpansion
}X_{A}\right\rangle \mathbf{I} \\ 
\left\langle X_{E^{\prime \prime }}X_{A}\right\rangle \mathbf{Z}%
\end{array}%
\right) =\left( 
\begin{array}{c}
\zeta \mathbf{I} \\ 
\eta \mathbf{Z}%
\end{array}%
\right),
\end{equation}%
where $\zeta =\sqrt{1-T}V_{A}$, $\eta =0$.

\textit{Reverse reconciliation}. The 3 dB loss limit on the transmission
line in the CV QKD \cite{Gros4} can be beaten with the reverse
reconciliation \cite{Gros3,Lu} or the post-selection \cite{Silber}. In
reverse reconciliation, Alice attempts to guess what was received by Bob
rather than Bob guessing what was sent by Alice \cite{Gros3}. Such a reverse
reconciliation protocol gives Alice an advantage over a potential
eavesdropper Eve. In reverse reconciliation, the secret key rate is 
\begin{equation}
K_{RR}=I(X_{A}\colon X_{B})-I(X_{B}\colon E),
\end{equation}%
where the mutual information between Alice and Bob $I(X_{A}:X_{B})$ is same
with what obtained above.

Eve's information is given by 
\begin{equation}
I(X_{B}\colon E)=S(E)-S(E|X_{B}).
\end{equation}

The conditional covariance matrix $\mathbf{\sigma }_{E\mid X_{B}}$, which
represents the covariance matrix of a system where one of the modes has been
measured by homodyne detection (in this case Bob), is given by \cite%
{Weedbrook2012,Eisert2002,Fiu2002} 
\begin{equation}
\mathbf{\sigma }_{E\mid X_{B}}=\mathbf{\sigma }_{E}-(V_{B})^{-1}\mathbf{D}%
\Pi \mathbf{D}^{T}.
\end{equation}%
Here $\mathbf{D}$ is the matrix describing the quantum correlations between
Eve' modes and Bob's mode, which is given by 
\begin{equation}
\mathbf{D=}\left( 
\begin{array}{c}
\left\langle X_{E%
%TCIMACRO{\U{b4}}%
%BeginExpansion
{\acute{}}%
%EndExpansion
}X_{B}\right\rangle \mathbf{I} \\ 
\left\langle X_{E^{\prime \prime }}X_{B}\right\rangle \mathbf{Z}%
\end{array}%
\right) =\left( 
\begin{array}{c}
\zeta ^{\prime }\mathbf{I} \\ 
\eta ^{\prime }\mathbf{Z}%
\end{array}%
\right),
\end{equation}%
where $\zeta ^{\prime }=\sqrt{T(1-T)}(W-V_{A})$, $\eta ^{\prime }=\sqrt{%
(1-T)(W^{2}-1)}$.

\begin{figure}[tbp]
\setlength{\belowcaptionskip}{-3pt} 
\centerline{
\includegraphics[width=70mm]{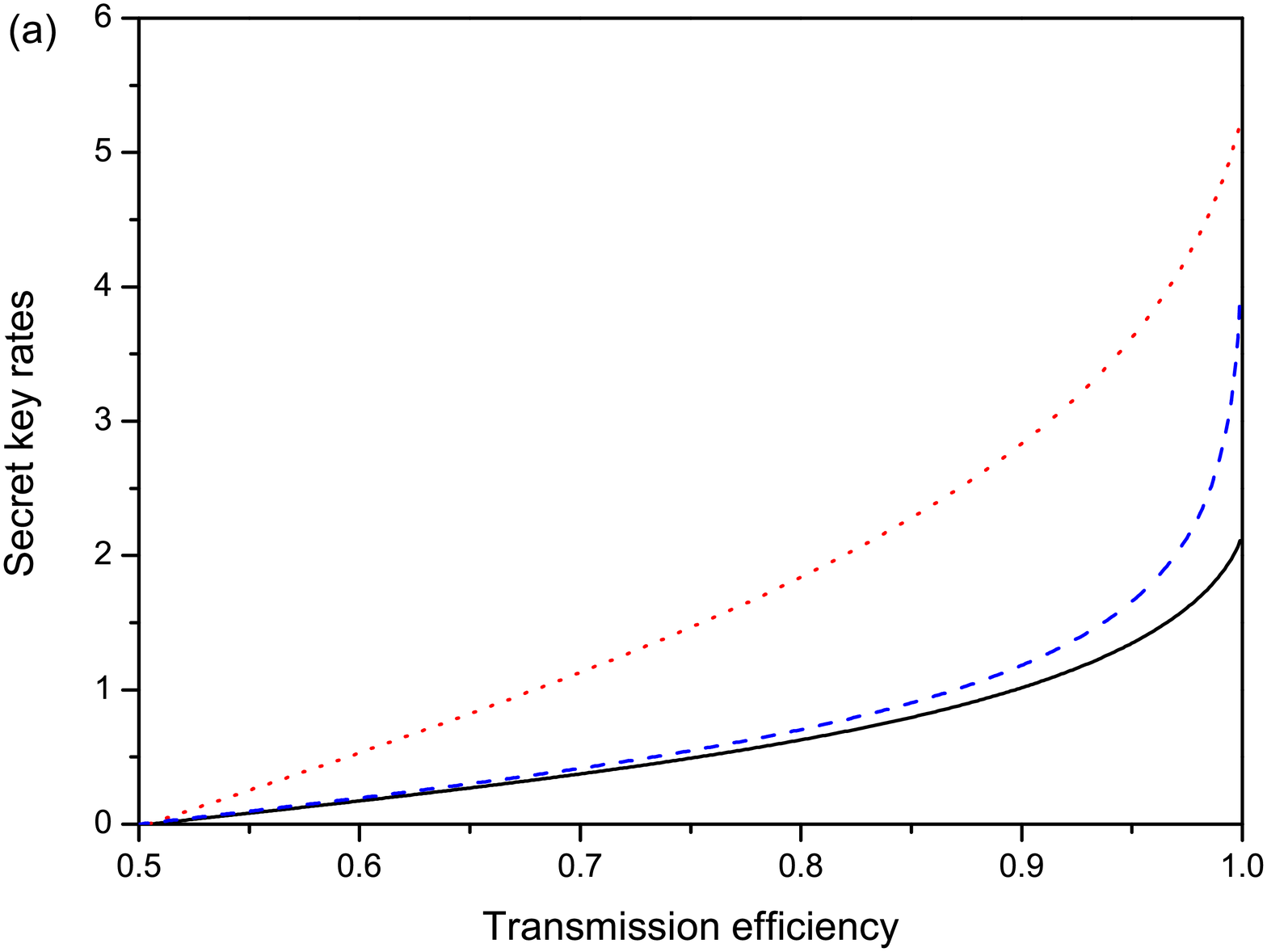}
}\includegraphics[width=70mm]{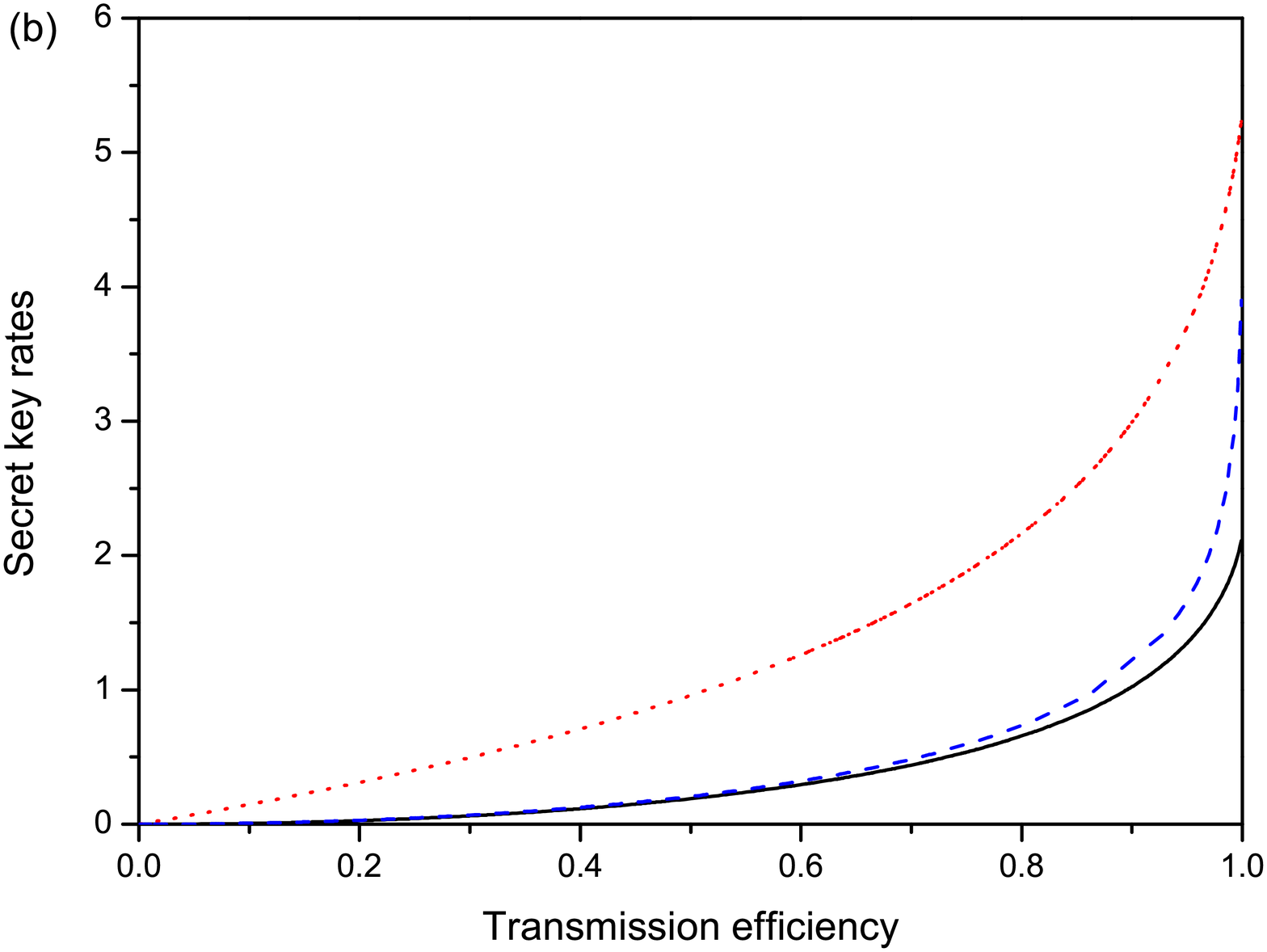}
\caption{Secret key rates for the CV QKD schemes with homodyne detection.
(a): the direct reconciliation, (b): the reverse reconciliation. Solid
(black) and Dashed (blue) lines are the secret key rates for the Gaussian
discord state with variance $V_{D}=40$ and 1000, respectively. Dotted (red)
line is the secret key rate for the entangled state with $V_{E}$=40. All
curves are plotted with excess noise W=1.}
\end{figure}

Figure 3 shows the secret key rate of the CV QKD scheme with homodyne
detection, (a) and (b) are corresponding to the direct and reverse
reconciliation, respectively. Solid (black) and Dashed (blue) lines are the
secret key rates for the Gaussian discord state with variance $V_{D}$=40
(typical experimental realistic modulation level \cite{Gros3}) and 1000,
respectively. Dotted (red) line is the secret key rate for the squeezed
state with variance $V_{E}$=40. All curves are plotted with excess noise
W=1. Comparing the solid and dotted lines in Fig. 3, it is obvious that
secret key rate for squeezed state is greater than that for Gaussian discord
state at the same energy level in both direct and reverse reconciliation.
Comparing solid and dashed lines, we find that the secret key rate is
increased with the increasing of the discording noise for the CV QKD with
the Gaussian discord state with homodyne detection in both direct and
reverse reconciliation.

\subsection{Heterodyne detection}

In the CV QKD scheme with heterodyne detection, Alice and Bob perform
heterodyne detection to measure the amplitude and phase quadratures of their
own beams simultaneously. Since heterodyning one of EPR entangled state is
equivalent to QKD with coherent state. In this section, we will compare the
Gaussian discord state QKD with no-switching coherent state QKD \cite%
{Weedbrook3}.

In heterodyne detection system, a vacuum mode $\hat{\nu}$ is mixed with the
optical mode $\hat{a}$ ($\hat{b}$) on a balanced beam-splitter and the
output modes are measured by two homodyne detectors respectively. The
amplitude quadrature measured by Alice and Bob are $\hat{X}_{A}^{M}=(\hat{X}%
_{a}+\hat{X}_{\nu })/\sqrt{2}$ and $\hat{X}_{B}^{M}=(\hat{X}_{B}+\hat{X}%
_{\nu })/\sqrt{2}$, respectively. The corresponding noise variance measured
by Alice and Bob are $V_{A}^{M}=(V_{A}+1)/2$ and $V_{B}^{M}=(V_{B}+1)/2$,
respectively.

Bob's conditional variance is given by $V_{B^{M}\mid A^{M}}=(V_{B\mid
A^{M}}+1)/2$, where 
\begin{equation}
V_{B\mid A^{M}}=V_{B}-\frac{\gamma ^{\prime 2}/2}{V_{A}^{M}}.
\end{equation}%
The mutual information between Alice and Bob are $I^{Het}(X_{A}\colon
X_{B})=\log _{2}[V_{B^{M}}/V_{B^{M}\mid A^{M}}]$, which is same for the
direct and reverse reconciliation.

\textit{Direct reconciliation. }In order to obtain $S(E|X_{B})$ we need to
calculate the symplectic spectrum of the conditional covariance matrix $%
\mathbf{\sigma }_{E\mid \hat{X}_{A},\hat{Y}_{A}}$, which represents the
covariance matrix of a system where two modes has been measured by
heterodyne detection (in this case Alice), is given by \cite%
{Weedbrook2012,Eisert2002,Fiu2002} 
\begin{equation}
\mathbf{\sigma }_{E\mid \hat{X}_{A},\hat{Y}_{A}}=\mathbf{\sigma }%
_{E}-(\Lambda )^{-1}\mathbf{D}(\Omega \mathbf{\sigma }_{A}\Omega ^{T}+%
\mathbf{I})\mathbf{D}^{T},
\end{equation}%
where $\Lambda =\det \mathbf{\sigma }_{A}+$Tr$\mathbf{\sigma }_{A}+1$, $%
\Omega \mathbf{\sigma }_{A}\Omega ^{T}+\mathbf{I=\sigma }_{A}+\mathbf{I}$,%
\textbf{\ }and $\mathbf{D}$ is given by eq. (18).

\textit{Reverse reconciliation}. The correlation matrix $\mathbf{\sigma }%
_{E\mid \hat{X}_{B},\hat{Y}_{B}}$, which represents the covariance matrix of
a system where two modes has been measured by heterodyne detection (in this
case Bob), is given by \cite{Weedbrook2012,Eisert2002,Fiu2002} 
\begin{equation}
\mathbf{\sigma }_{E\mid \hat{X}_{B},\hat{Y}_{B}}=\mathbf{\sigma }%
_{E}-(\Lambda ^{\prime })^{-1}\mathbf{D}(\Omega \mathbf{\sigma }_{B}\Omega
^{T}+\mathbf{I})\mathbf{D}^{T},
\end{equation}%
where $\Lambda ^{\prime }=\det \mathbf{\sigma }_{B}+$Tr$\mathbf{\sigma }%
_{B}+1$, $\Omega \mathbf{\sigma }_{B}\Omega ^{T}+\mathbf{I=\sigma }_{B}+%
\mathbf{I}$, and the matrix $\mathbf{D}$ is same with eq. (22), which
describing the quantum correlations between Eve' modes and Bob's mode.

\begin{figure}[tbp]
\setlength{\belowcaptionskip}{-3pt} 
\centerline{
\includegraphics[width=70mm]{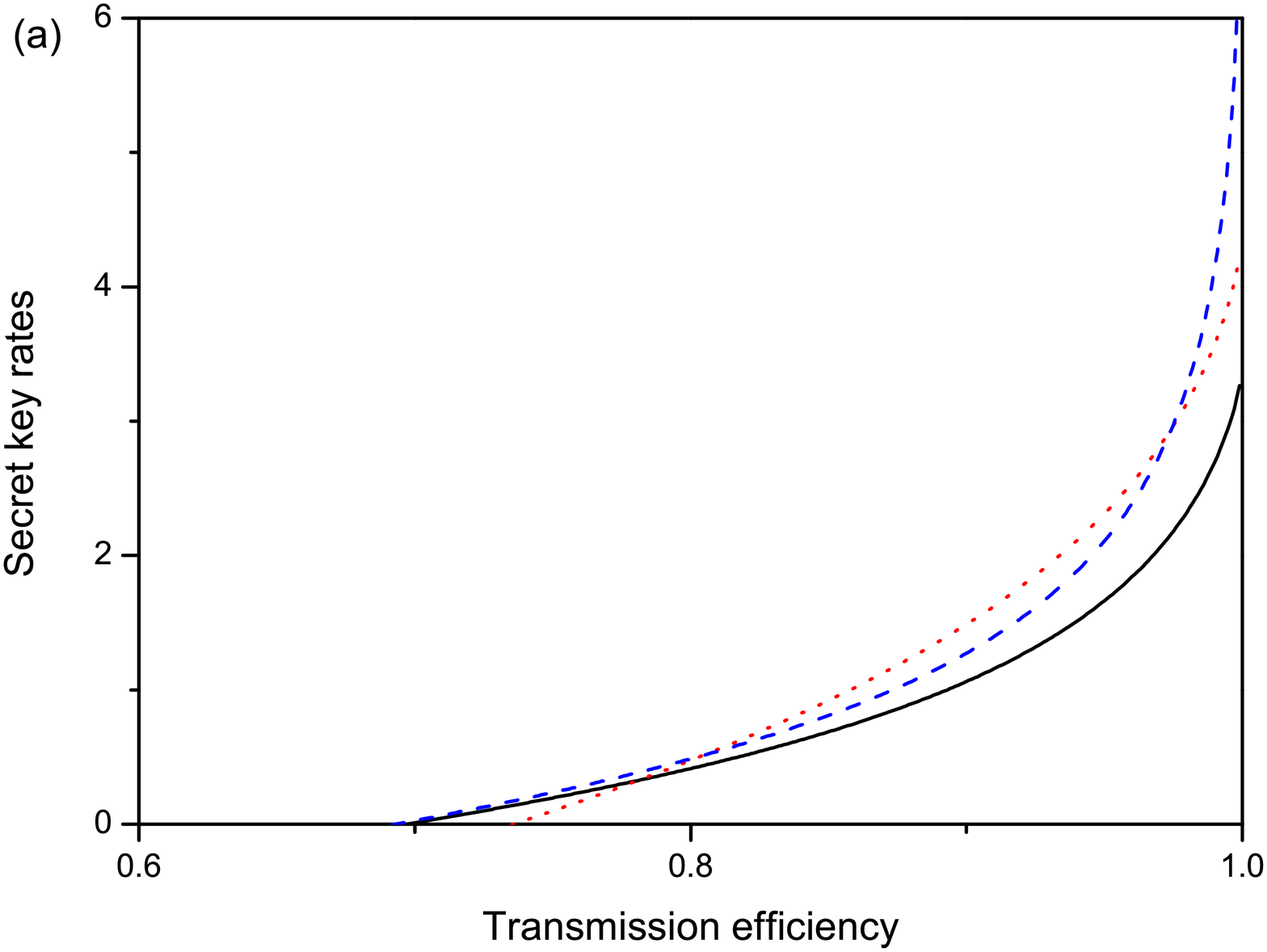}
}\includegraphics[width=70mm]{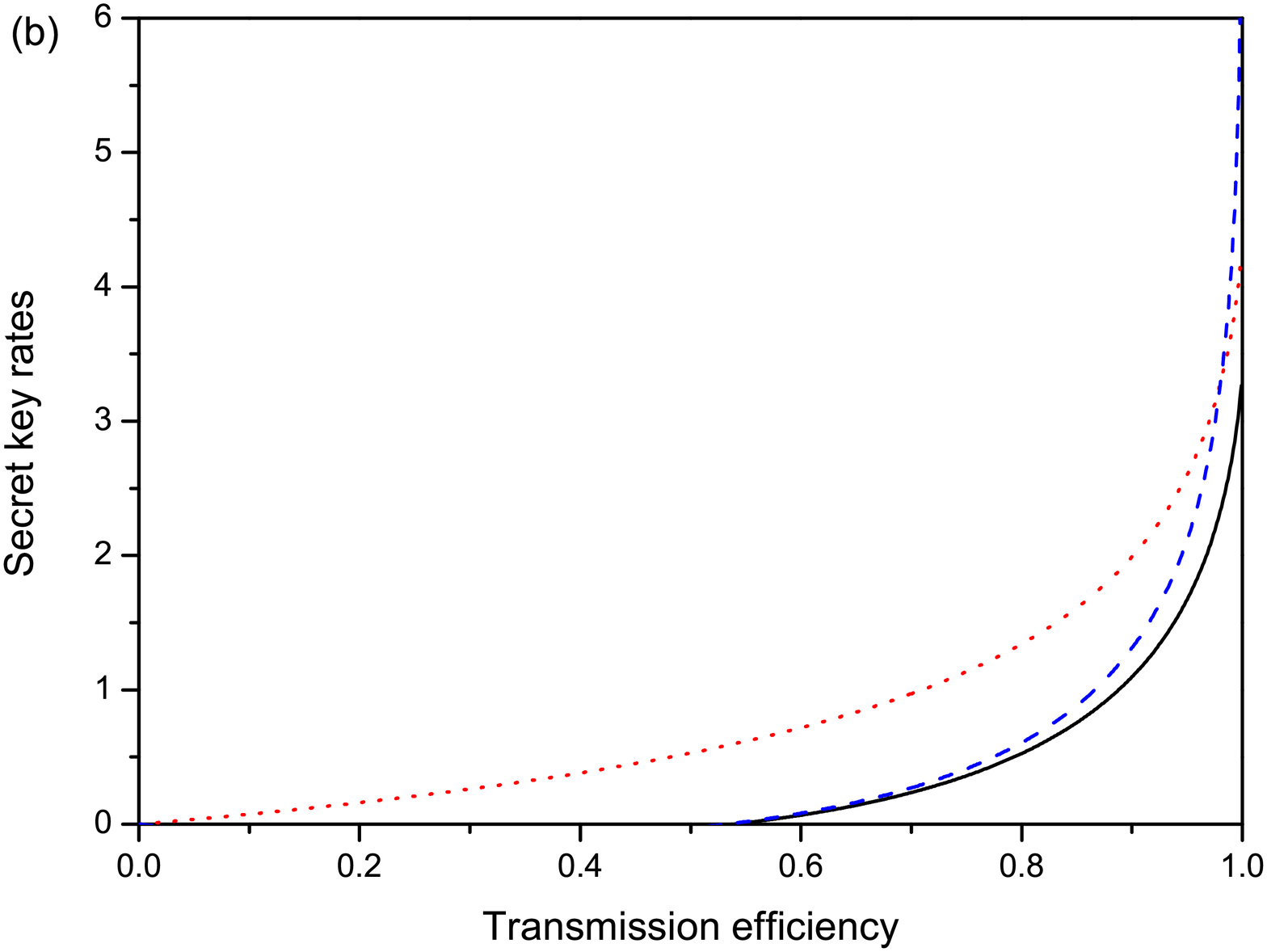}
\caption{Secret key rates for the CV QKD schemes with heterodyne detection.
(a): the direct reconciliation, (b): the reverse reconciliation. Solid
(black) and Dashed (blue) lines are the secret key rates for the Gaussian
discord state with $V_{D}$=40 and 1000, respectively. Dotted (red) line is
the secret key rate for the entangled state with $V_{E}$=40. All curves are
plotted with excess noise W=1.}
\end{figure}

Figure 4 shows the secret key rates for the CV QKD schemes with heterodyne
detection, (a) and (b) are for the direct and reverse reconciliation,
respectively. Solid (black) and Dashed (blue) lines are the secret key rates
for the Gaussian discord state with $V_{D}=40$ and 1000, respectively.
Dotted (red) line is the secret key rate for the entangled state with $%
V_{E}=40$. All curves are plotted with excess noise $W=1$. In Fig. 4(a),
comparing solid and dotted lines, we find that secret key can be distilled
for the Gaussian discord state at lower transmission efficiency than that
for coherent state with heterodyne detection. When $T>0.78$, secret key rate
for coherent state is still higher than that for the Gaussian discord state
with heterodyne detection. In Fig. 4(b), comparing solid and dotted lines,
it is obvious that no-switching coherent state QKD offers higher secret key
rate and longer transmission distance than that the Gaussian discord state
QKD. We also noticed that no secret key can be distilled for the Gaussian
discord state at lower transmission efficiency ($T<0.55$) with reverse
reconciliation, which is different from coherent state QKD. Comparing solid
and dashed lines in Fig. 4(a) and (b), respectively, we find that secret key
rate is increased with increasing of the discording noise for both direct
and reverse reconciliation in CV QKD with the Gaussian quantum discord
state, which is same with the result of homodyne detection.

\section{Dependence of secret key rate on quantum discord}

\begin{figure}[tbp]
\setlength{\belowcaptionskip}{-3pt} 
\centerline{
\includegraphics[width=80mm]{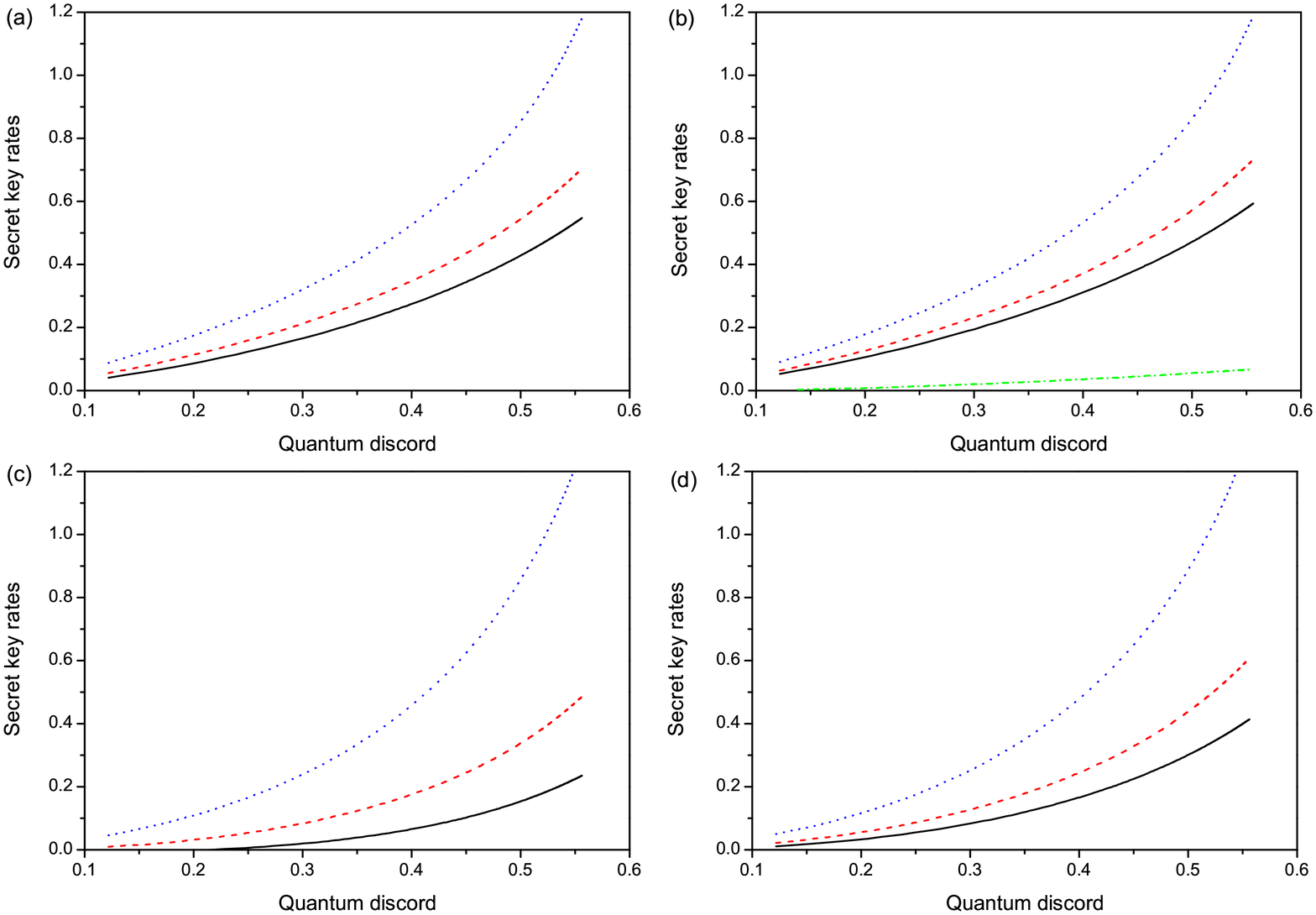}
}
\caption{The dependence of secret key rates on quantum discord for the CV
QKD schemes with the Gaussian discord state. (a) and (b): the direct and
reverse reconciliation for homodyne detection, respectively. (c) and (d):
the direct and reverse reconciliation for heterodyne detection,
respectively. Solid (black), Dashed (red) and Dotted (blue) lines are the
secret key rates for the Gaussian discord state with transmission efficiency
of 0.75, 0.8 and 0.9, respectively. Dash-dotted (green) line in (b) is the
secret key rate for the Gaussian discord state with transmission efficiency
of 0.3. All curves are plotted with excess noise W=1, $V_{D}$ $\in $ $%
[1,1000]$.}
\end{figure}

As shown in Fig. 5, the dependence of secret key rate for the Gaussian
discord state on quantum discord are investigated at different transmission
efficiency with input variance $V_{D}\in \lbrack 1,1000]$. Fig. 5(a) and (b)
are the case of direct and reverse reconciliation for homodyne detection,
respectively. Fig. 5(c) and (d) are the case of direct and reverse
reconciliation for heterodyne detection, respectively. It is obvious that
secret key rate is monotonically increased with the increasing of quantum
discord. Solid (black), Dashed (red) and Dotted (blue) lines are the secret
key rates for the Gaussian discord state with transmission efficiency of
0.75, 0.8 and 0.9, respectively. Dash-dotted (green) line in Fig. 5(b) is
the secret key rate for the Gaussian discord state with transmission
efficiency of 0.3, which means that secret key can be distilled when $T<0.5$
in reverse reconciliation for homodyne detection. Comparing these traces, we
find that secret key rate is increased with the increasing of transmission
efficiency, which is same with the result in Fig. 3 and 4. Most of the
secret key rates start from $D_{AB}=0.12$, since 0.12 is the smallest
quantum discord with $V_{D}=1$ as shown in Fig. 2(a). When $T=0.75$ (solid
line) in Fig. 5(c), secret key can be distilled when $D_{AB}>0.22$.

\section{Conclusion}

In this paper, by considering CV QKD with a two-mode Gaussian discord state,
which has only quantum correlation and without entanglement, we show that
secret key can be distilled against entangling cloner attack. In CV QKD with
the Gaussian discord state,the secret key rate is increased with increasing
of quantum discord in both homodyne and heterodyne detection schemes with
direct and reverse reconciliation. By comparing the secret key rates of CV
QKD schemes with the Gaussian discord state, squeezed state and coherent
state, we find that squeezed state and coherent state offer higher secret
key rate than the Gaussian discord state at the same energy level for both
direct and reverse reconciliation. This is a natural result since Gaussian
discord of the Gaussian discord state ($0\leq D\leq 1$) is smaller than that
of EPR entangled state ($D>1$). This work provides a possible application of
Gaussian quantum discord.

\section*{Acknowledgments}

The author thanks for helpful discussion with Prof. Changde Xie, Kunchi
Peng, Jing Zhang and Xiaojun Jia. This research was supported by the
National Basic Research Program of China (Grant No. 2010CB923103), NSFC
(Grant Nos. 11174188, 61121064), Shanxi Scholarship Council of China (Grant
No. 2012-010) and OIT.

\end{document}